\documentclass{aa}

\usepackage{graphicx}

\begin{document}

   \title{SuperWASP discovery and SALT confirmation of a semi-detached eclipsing binary that
    contains a $\delta$~Scuti star}

   \titlerunning{$\delta$ Scuti star in an eclipsing binary}

   \author{A.J. Norton\inst{1} \and M.E. Lohr\inst{1} \and B. Smalley\inst{2}
   \and P.J. Wheatley\inst{3} \and R.G. West\inst{3} }

    \authorrunning{Norton et al.}

   \institute{Department of Physical Sciences, The Open University, Walton Hall,
   Milton Keynes MK7 6AA, U.K.
   \and
   Astrophysics Group, Keele University, Staffordshire ST5 5BG, U.K.
   \and
   Department of Physics, University of Warwick, Coventry CV4 7AL, U.K.
   }

   \date{Received July 3, 2015; accepted January 12, 2016}


  \abstract
 {}
   {We searched the SuperWASP archive for objects that display multiply periodic photometric variations.}
   {Specifically we sought evidence for eclipsing binary stars that display a further non-harmonically related signal in their power spectra.}
   {The object \object{1SWASP~J050634.16--353648.4} has been identified as a relatively bright ($V \sim 11.5$) semi-detached eclipsing binary with a 5.104~d orbital period that displays coherent pulsations with a semi-amplitude of 65~mmag at a
   frequency of 13.45~d$^{-1}$. Follow-up radial velocity spectroscopy with the Southern African Large Telescope confirmed the binary nature of the system. Using the {\sc phoebe} code to model the radial velocity curve with the SuperWASP photometry enabled parameters of both stellar components to be determined. This yielded a primary (pulsating) star with a mass of $1.73 \pm 0.11$~M$_{\odot}$ and a radius of $2.41 \pm 0.06$~R$_{\odot}$, as well as a Roche-lobe filling secondary star with a mass of $0.41 \pm 0.03$~M$_{\odot}$ and a radius of $4.21 \pm 0.11$~R$_{\odot}$.}
   {\object{1SWASP~J050634.16--353648.4} is therefore a bright $\delta$ Sct pulsator in a semi-detached eclipsing binary with one of the largest pulsation amplitudes of any such system known. The pulsation constant indicates that the mode is likely a first overtone radial pulsation.}


   \maketitle
%

\section{Introduction}

SuperWASP (Pollacco et al. 2006) is the world's leading ground-based survey for transiting exoplanets. As a spin-off from its main science, it provides a long baseline, high cadence, photometric survey of bright stars across the entire sky, away from the Galactic plane. Operating since 2004, the archive contains over 2000 nights of data, comprising light curves of over 30 million stars in the magnitude range $\sim 8 - 15$, with typically 10000 observations per object. It therefore provides a unique resource for studies of stellar variability (e.g. Norton et al. 2007; Smalley et al. 2011, 2014; Delorme et al. 2011; Lohr et al. 2013, 2014a, 2014b, 2015a, 2015b; Holdsworth et al. 2014). SuperWASP data is pseudo $V$-band, with 1~count~s$^{-1}$ equivalent to $V \sim 15$. Since the individual SuperWASP telescopes have a small aperture (11~cm), the data are subject to significant noise and uncertainties, but the long baseline and frequent observations are able to compensate for these limitations to some extent.

Stellar systems containing two or more underlying `clocks' are usually interesting objects to study and can provide critical insight into various aspects of stellar astrophysics. Zhou (2014) provides a catalogue of `oscillating binaries' of various types that includes 262  pulsating components in eclipsing binaries (96 of which contain $\delta$~Sct stars) plus a further 201 in spectroscopic binaries (including 13 $\delta$~Sct stars) and 23 in visual binaries (including one $\delta$~Sct star). They list a number of open questions that may be addressed by such systems, including the influence of binarity on pulsations and the link between them. More specifically, Liakos et al. (2012), updated in Liakos \& Niarchos (2015), recently published a summary of 107 eclipsing binary systems that include a pulsating $\delta$ Sct component. They emphasise that single $\delta$ Sct stars exhibit different evolutionary behaviour to their counterparts in binary systems, which is likely related to mass transfer and tidal distortions occurring in the binaries. Since only around 100 $\delta$~Sct stars are known in eclipsing binaries, any newly identified system is noteworthy, particularly if it is bright enough for follow-up using modest instruments and if it displays a behaviour at the extremes of the rest of the sample.

\section{SuperWASP data}

To search for further examples of pulsating stars in binary systems, we investigated the SuperWASP archive. Previously we  carried out a period-search across the entire SuperWASP data set (for details of the method employed, see Norton et al. 2007) and preliminary catalogues of eclipsing binaries and pulsating stars are presented in the PhD thesis by Payne (2012). Owing to the many systematic sources of noise in the data, the period-searching carried out typically results in multiple apparently significant periods being allocated to each object, many of which turn out to be spurious on individual investigation, since the spurious periods are usually close to harmonics of one day or one month and may be eliminated.

The approach taken here was to perform an automated search through all the potential periods identified in each SuperWASP object, picking out systems that had more than one period with high significance and for which the periods identified were \emph{not} harmonically related to one another. Care was also taken to exclude periods related to each other by daily aliases, or harmonics of those aliases. Having done this, systems were identified which contained pairs of significant periods that were unrelated to each other, and thus likely represented two independent clocks in the system.

\begin{figure}[ht]
   \begin{center}
   \includegraphics[scale=0.4,angle=0]{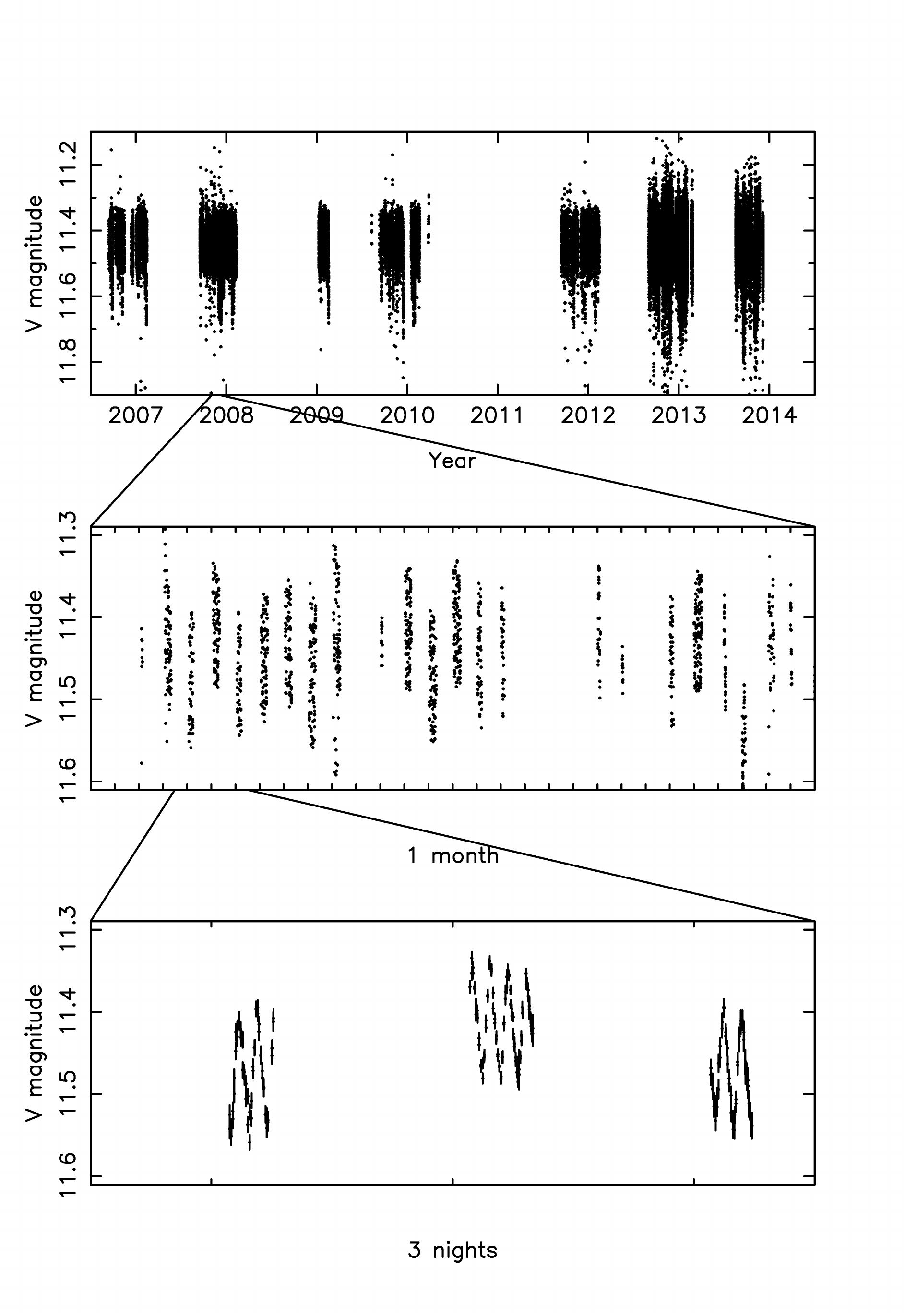}
      \caption{Light curve of the SuperWASP data of 1SWASP~J050634.16-353648.4, which contains over 56000 data points, spanning seven years. The upper panel shows the complete dataset (with no error bars as they would obscure the data at this scale), illustrating the long term coverage; the middle panel contains a representative month's worth of data from late 2007 (again with no error bars), illustrating that regular observations occur on most nights during an observing season and showing hints of the longer-term modulation owing to the eclipsing binary; the lower panel presents three nights of data and shows the observed pulsation, including error bars.}
   \label{lcurve}
   \end{center}
   \end{figure}

A candidate eclipsing binary that contains a pulsating star was identified and is the subject of this research. The coordinates of \object{1SWASP~J050634.16--353648.4} (= TYC~7053--566--1) are reflected in its name, which is derived from its location in the USNO-B1 catalogue. This is an anonymous 11th magnitude star ($V =11.514, B=11.767$) that is clearly isolated both in the SuperWASP images and sky survey images. The SuperWASP photometric data comprise 56093 data points spanning over seven years from 2006 September 14 to 2013 December 6 (Fig.~\ref{lcurve}).

\begin{figure}[ht]
   \begin{center}
   \includegraphics[scale=0.4,angle=0]{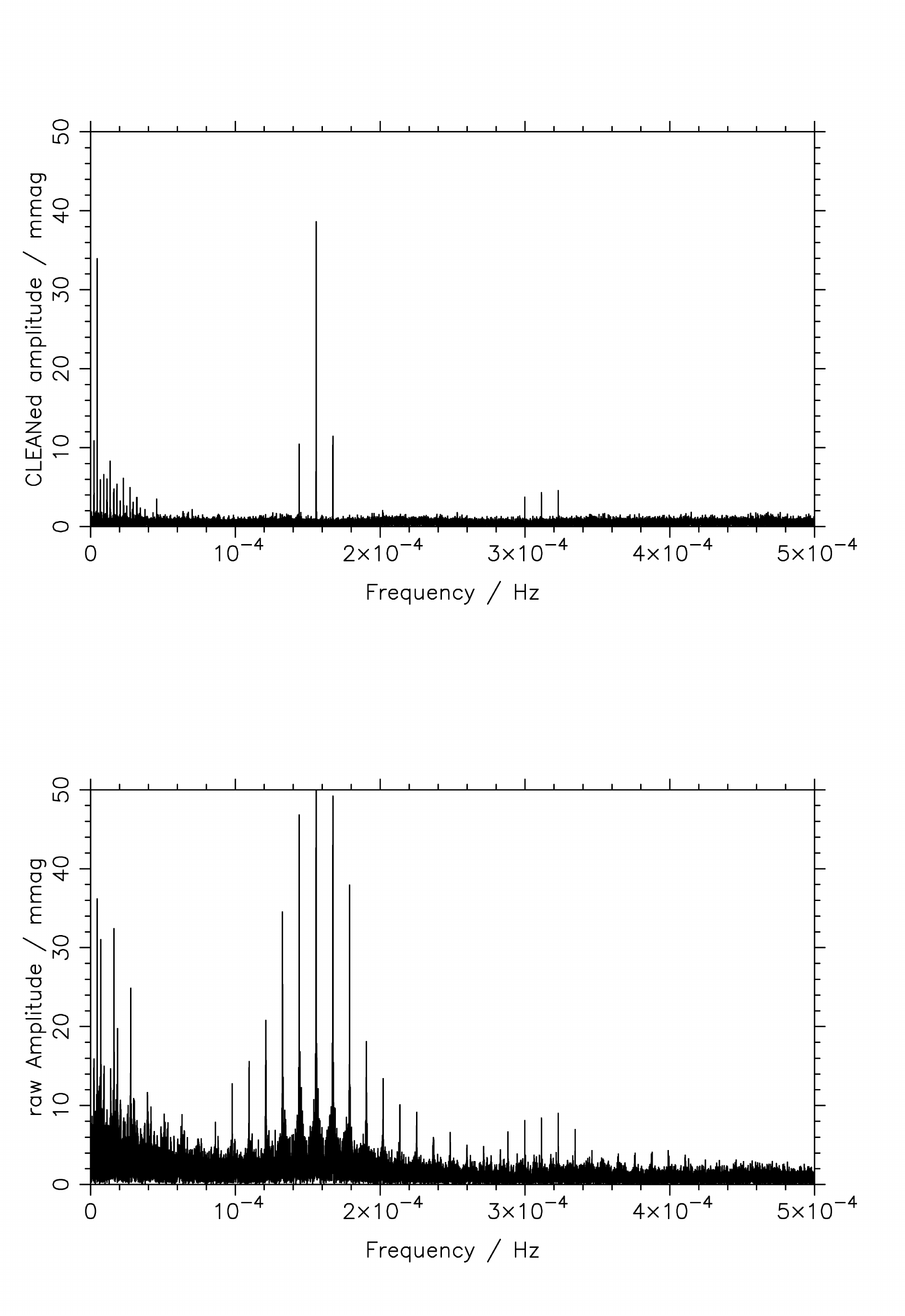}
      \caption{Power spectrum of the SuperWASP data of \object{1SWASP~J050634.16--353648.4}. The lower panel is the raw power spectrum, the upper panel is the {\sc clean}ed power spectrum, both with amplitudes in mmag. Strong peaks at frequencies of $1.557 \times 10^{-4}$~Hz (or 13.45~d$^{-1}$) and $2.268 \times 10^{-6}$~Hz (or
      a period around 5.104~d) are seen.}
   \label{pspec}
   \end{center}
   \end{figure}

We carried out a power spectrum analysis of the light curve using an implementation of the {\sc clean} algorithm (H\"{o}gbom 1974; Roberts, Leh\'{a}r \& Dreher 1987) by Lehto (1993). This version utilises variable gain, thereby rendering it less susceptible to falsely cleaning out a genuine periodic signal when the data are particularly noisy. The resulting power spectrum shows two distinct sets of signals (Fig.~\ref{pspec}). The first comprises a strong peak, which corresponds to a frequency of $1.5570803(1) \times 10^{-4}$~Hz along with its first harmonic. Sidebands to each, at $\pm 1$ sidereal day, remain even after moderate {\sc clean}ing.
There is a second set of frequency components with a primary period around
5.104~d, although the first harmonic at half the period is the stronger peak, and all harmonics up to $14 \times$ the fundamental frequency can be seen.

\begin{figure}[ht]
   \begin{center}
   \includegraphics[scale=0.5,angle=0]{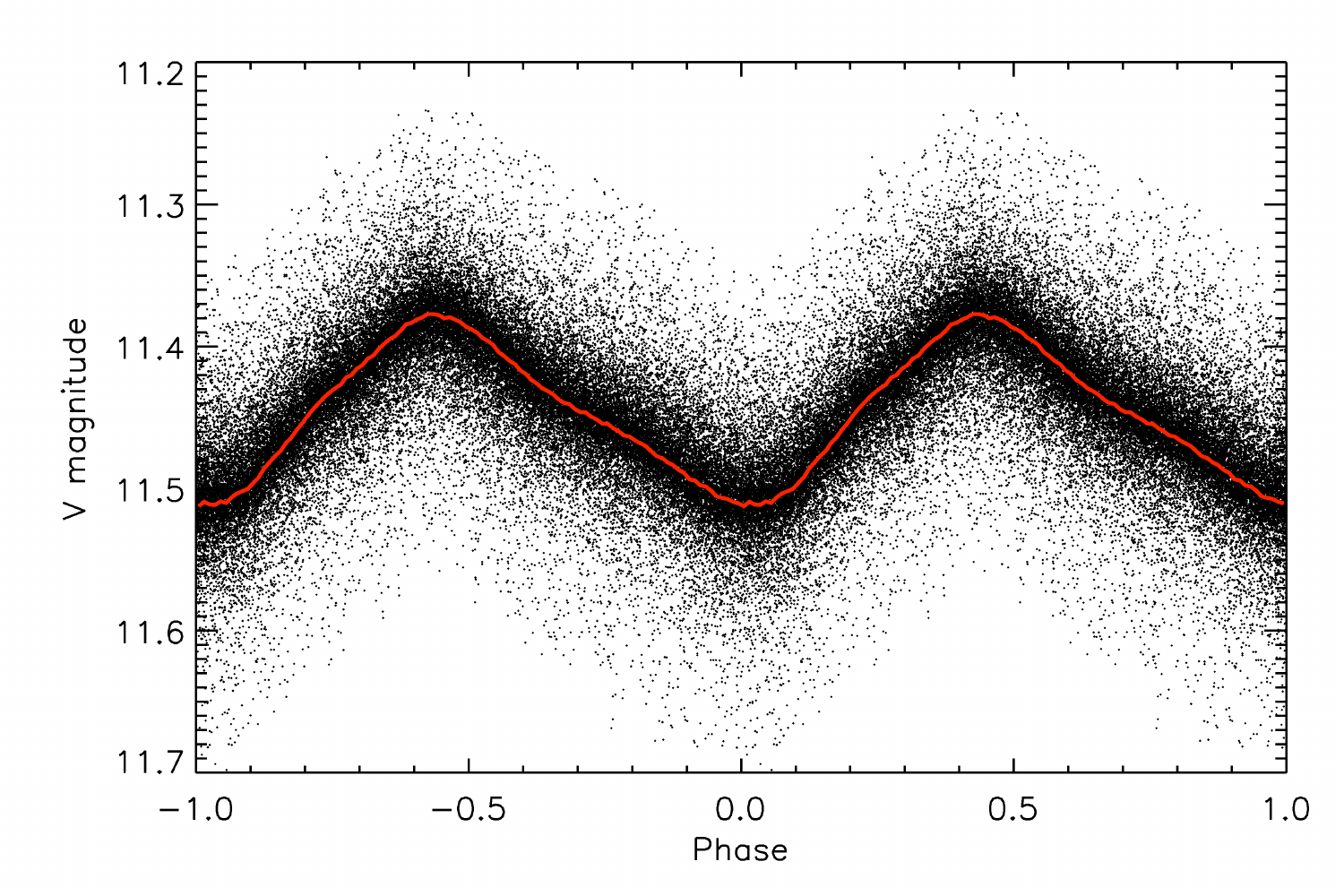}
   \includegraphics[scale=0.5,angle=0]{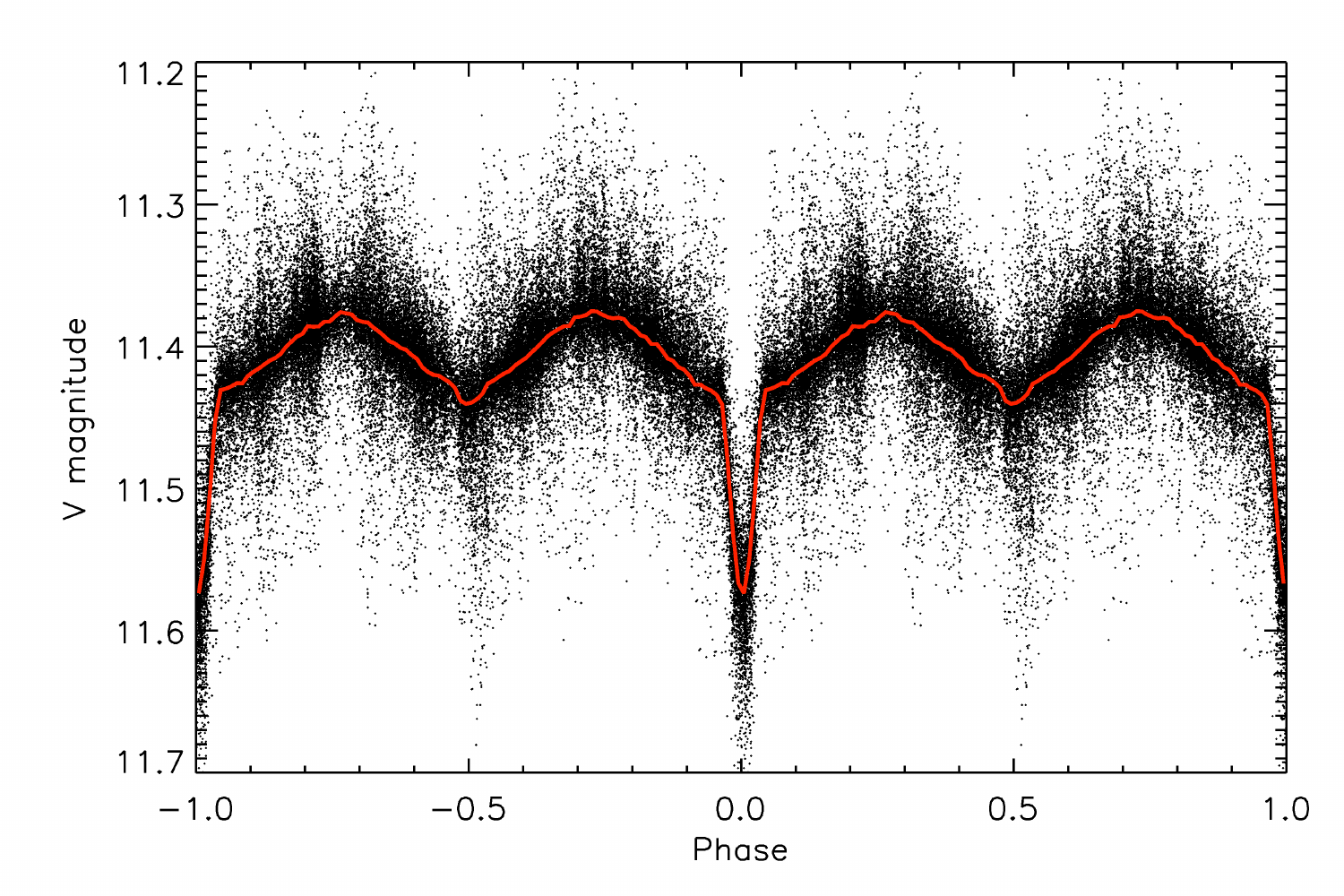}
      \caption{The SuperWASP data of \object{1SWASP~J050634.16--353648.4,} folded at the pulsation period of 1.78396567~h (top) and the binary orbital period of 5.104238~d (bottom). In each case, the mean-binned light curve folded at the other period has been removed from the data. Phase zero corresponds to the minimum of the pulse in the upper panel and the primary eclipse in the lower panel; the red solid line shows the binned profile in each case.}
   \label{fold}
   \end{center}
   \end{figure}

 Folding the data at the shorter period reveals a quasi-sinusoidal pulsation, but with a significant scatter that is caused by the variability at the other, longer period. To separate the two signals, a template pulsational light curve was first obtained as follows. The folded data were divided into 100 phase bins. The optimally-weighted average flux in each bin was calculated as the inverse variance-weighted mean of all the fluxes in the bin (e.g. Horne 2009). A spline curve was then interpolated through these binned average points to create a template pulse profile. Copies of this template were then subtracted from the original light curve to leave a residual light curve that we observed as being characteristic of an eclipsing binary. We then carried out a phase dispersion minimization analysis (e.g. Schwarzenberg-Czerny 1997) on the resulting light curve and identified the longer period more precisely as 5.104238(5)~d.
A template for the eclipsing binary light curve was then constructed from these residual phase-binned data using the optimally-weighted average flux for each bin, as before, and then fitting a spline curve to the eclipse profile. A cleaner light curve for the pulsator was then obtained by subtracting this eclipsing binary template from the full original data set. The whole procedure of creating optimally-weighted average profiles for each of the two modulations and subtracting them in turn from the original light curve was iterated a few times to ensure convergence on representative light curves and periods. The procedure was halted when the resulting folded light curves were indistinguishable from those of the previous iteration, within the data uncertainties. The final separated data, folded at the identified periods, are shown in Fig.~\ref{fold} with the binned profile overlaid as a red line in each case.

 The semi-amplitude of the pulsation is $65\pm7$~mmag, and the primary eclipses have a depth of $195\pm7$~mmag. However, since each of these  amplitudes is measured from a folded light curve that has had the other modulation removed from it, there may be some systematic uncertainty in these values.

The times of primary minima for the eclipses may be represented by
\[ T_0(\mathrm{HJD/UTC}) = 2456992\fd 484(9) + 5\fd 104238(5) E, \]
where the figures in parentheses represent the uncertainties in the last digits.
Over the seven year baseline that is available, this displays no measurable evidence for eclipse-timing variations that might indicate a secular period change or the presence of a third body in the system. The primary and secondary eclipses are separated by precisely half the orbital cycle, indicating an eccentricity, $e$, of zero.

\section{Spectroscopic data}

\begin{figure}[ht]
   \begin{center}
   \includegraphics[scale=0.5,angle=0]{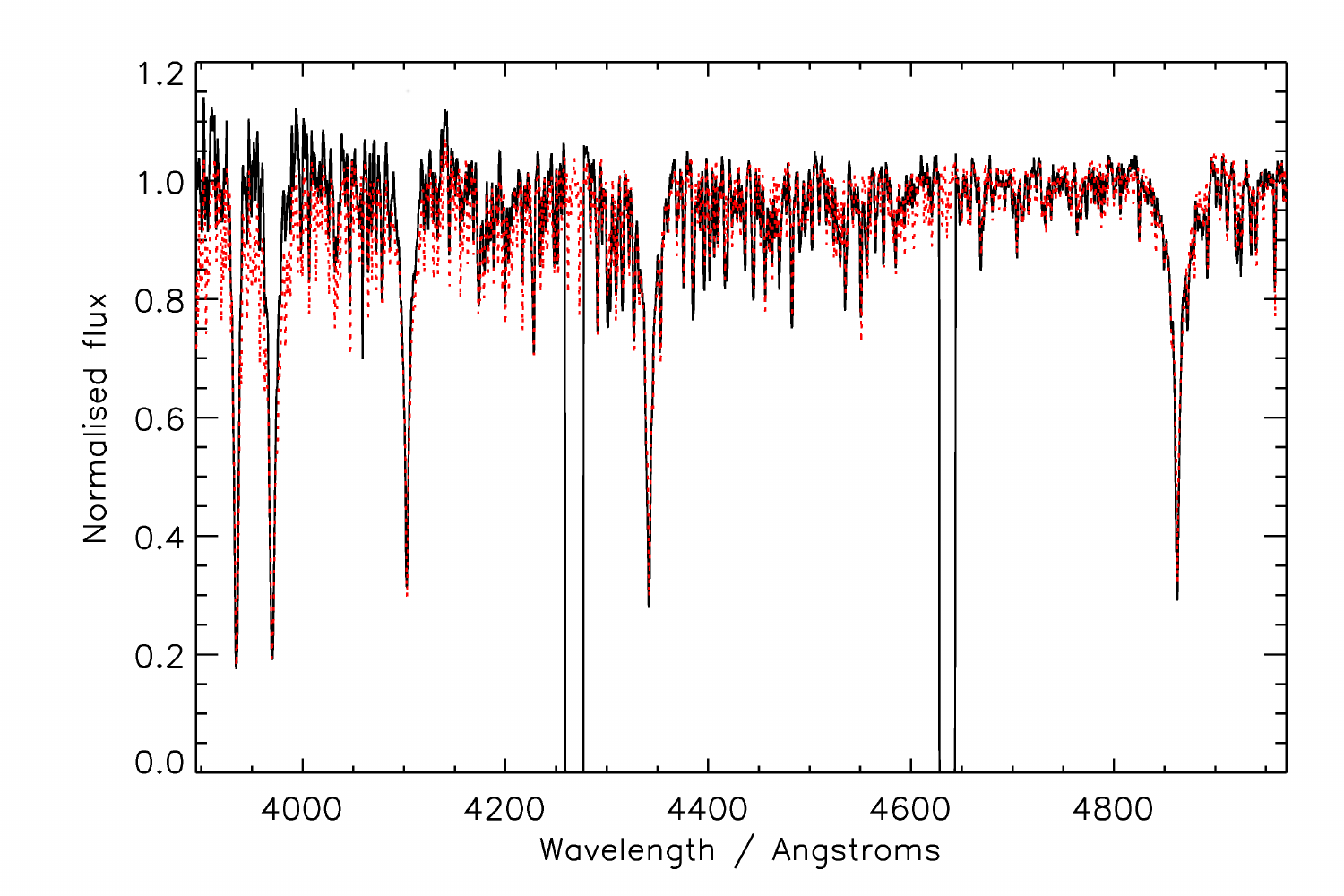}
      \caption{A normalised SALT spectrum of \object{1SWASP~J050634.16--353648.4}, taken close to orbital phase zero. The red line shows the best-matching {\sc phoenix} template, which has $T = 7000$~K and $\log g= 4.0$. Most of the deep absorption lines are Balmer lines from H$\beta$ to H$\epsilon$. }
   \label{spectrum}
   \end{center}
   \end{figure}

To confirm the nature of the system, we obtained a set of fifteen spectra of \object{1SWASP~J050634.16--353648.4} via queue-scheduled observing on ten visits between December 2014 and February 2015 using the Southern African Large Telescope (SALT: Buckley et al. 2006). Observations were obtained using the Robert Stobie Spectrograph (RSS: Burgh et al. 2003) with the pg2300 grating, giving a dispersion of 0.17\AA \ per pixel in the wavelength ranges 3896--4257\AA, 4278--4623\AA, and 4642--4964\AA \ across the three chips of the detector. An observation log is given in Table~1. The spectra were reduced using standard {\sc iraf} tools (Tody 1986, 1993) and an example spectrum is shown in Fig.~\ref{spectrum}.

\begin{table*}
\caption{Log of SALT observations. Exposure times were between 100s and 120s for each spectrum. SNR indicates the signal-to-noise ratio around 4800~\AA. RV$_1$ and RV$_2$ are the radial velocities obtained via cross-correlation; RV$_{1*}$ and  RV$_{2*}$ are those obtained from the disentangled spectra.}
\begin{tabular}{llccrrrr} \hline
Date & Start (UT)  & SNR & Orbital phase & RV$_1$/ km~s$^{-1}$ & RV$_2$ / km~s$^{-1}$  & RV$_{1*}$/ km~s$^{-1}$ & RV$_{2*}$ / km~s$^{-1}$ \\ \hline
2014-12-01  & 01:46:43  &  98  & 0.0184 &  \multicolumn{2}{c}{$-19.7 \pm 2.0$} & $-32.1\pm2.9$ & $-22.6\pm3.6$ \\
2014-12-01      & 01:48:36      &  109 & 0.0186 &  \multicolumn{2}{c}{$-16.2 \pm 2.1$} & $-29.1\pm1.9$ & $-22.0\pm7.4$ \\
2014-12-06      & 01:29:19      &  102 & 0.9956 &  \multicolumn{2}{c}{$-27.2 \pm 2.1$} & $-33.5\pm4.7$ & $-45.6\pm9.8$ \\
2014-12-06      & 01:31:12      &  102 & 0.9958 &  \multicolumn{2}{c}{$-28.9 \pm 2.0$} & $-35.2\pm3.9$ & $-50.4\pm14.0$ \\
2014-12-07      & 01:41:40      &  118 & 0.1932 & $-62.5 \pm 3.0$ & $+66.8 \pm 21.9$ & $-69.9\pm5.6$ & $+71.1\pm6.8$ \\
2014-12-07      & 01:43:33      &  112 & 0.1934 & $-54.4 \pm 2.8$ & $+83.3 \pm 27.0$ & $-63.1\pm4.6$ & $+75.0\pm5.0$ \\
2014-12-12      & 01:22:18      &  138 & 0.1701 & $-57.6 \pm 3.3$ & $+79.3 \pm 26.0$ & $-66.7\pm11.9$ & $+60.4\pm16.6$ \\
2014-12-12      & 01:24:15      &  145 & 0.1704 & $-56.5 \pm 3.5$ & $+79.5 \pm 29.6$ & $-67.2\pm11.7$ & $+57.8\pm18.2$ \\
2014-12-15      & 19:16:51      &  114 & 0.9040 & $-18.2 \pm 2.9$ & $-105.0 \pm 19.1$ & $-22.5\pm2.6$ & $-107.4\pm5.1$ \\
2014-12-15      & 19:18:48      &  112 & 0.9043 & $-18.5 \pm 3.0$ & $-99.1 \pm 20.2$ & $-22.4\pm2.2$ & $-106.9\pm5.7$ \\
2015-01-01      & 23:32:13      &  127 & 0.2693 & $-62.6 \pm 2.5$ & $+83.5 \pm 25.3$ & $-62.7\pm1.8$ & $+93.4\pm7.7$ \\
2015-01-04      & 23:36:56      &  142 & 0.8597 & $-19.5 \pm 3.0$ & $-136.4 \pm 21.3$ & $-18.3\pm1.7$ & $-122.5\pm13.9$ \\
2015-01-06      & 23:25:25      &  126 & 0.2479 & $-61.3 \pm 3.0$ & $+87.4 \pm 24.6$ & $-59.8\pm2.5$ & $+93.6\pm9.9$ \\
2015-01-21      & 23:00:41      &  62  & 0.1832 & $-39.8 \pm 3.1$ & $+87.3 \pm 26.0$ & $-38.7\pm4.0$ & $+79.3\pm9.8$ \\
2015-02-04      & 21:58:29  &  96  & 0.9174 & $-13.9 \pm 3.2$ & $-106.5 \pm 34.2$ & $-5.2\pm4.9$ & $-84.8\pm8.0$ \\ \hline
\end{tabular}
\end{table*}

The spectra revealed two sets of lines that correspond to the two components of the system. Radial velocities for each component were measured from all spectra (except those close to phase zero where only single blended lines are seen) by cross-correlation with a best-matching template from the {\sc phoenix} synthetic stellar library (Husser et al. 2013). Orbital phases and radial velocities from cross-correlating each spectrum are listed in Table~1. The template used had an effective temperature of 7000~K and solar abundances, which correspond to spectral type around F1. In the wavelength region covered by the RSS in these observations, the spectrum of \object{1SWASP~J050634.16--353648.4} is dominated by the light from the primary star, so this is appropriate.

\section{Modelling the system}

We modelled \object{1SWASP~J050634.16--353648.4} by a combination of spectral disentangling using {\sc korel} (Hadrava 2004, 2012)  and the {\sc phoebe} interface (Pr\u{s}a \& Zwitter 2005) to the Wilson-Devinney code (Wilson \& Devinney 1971).

\begin{figure}[ht]
   \begin{center}
   \includegraphics[scale=0.6,angle=0]{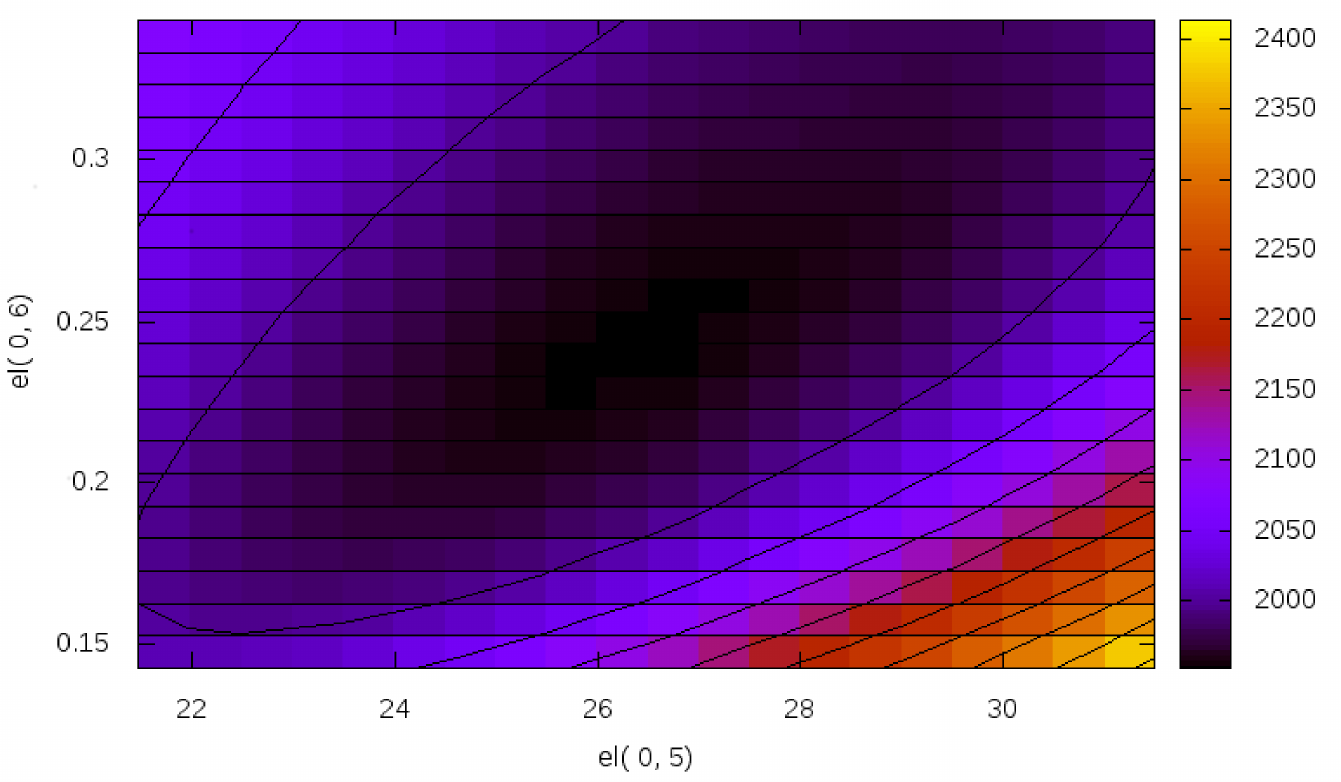}
      \caption{Joint confidence levels for the mass ratio $q$ (vertical axis) and primary semi-amplitude $K_1$ (horizontal axis) resulting from {\sc korel}. The colour scale indicates the relative goodness of fit, with darker blue corresponding to a better fit.}
   \label{qvsk1}
   \end{center}
   \end{figure}

To work successfully, {\sc korel} requires that the margins of the wavelength region in the analysed spectra  contain only continuum  with no spectral lines. There was no such extended region at the extremes of our spectra so we used seven separate wavelength ranges, each containing several lines with continuum on either side. The seven short spectral regions were independently analysed to find $K_1$ (the semi-amplitude of the primary component) and $q$ (the mass ratio $M_2/M_1$), using the estimates found from cross-correlation as a starting point.  All converged on very similar values, giving means of $q=0.236\pm0.004$ and $K_1=28.1 \pm 0.5$~km~s$^{-1}$. Figure~\ref{qvsk1} shows the shape of the parameter space around this location and the limited correlation between the two converged parameters.  From these, $K_2=118.8$~km~s$^{-1}$ may be derived, implying $a \sin i=14.8 \pm 0.3~\mathrm{R}_{\odot}$.  Radial velocities were also obtained for each exposure from the best superposition of the decomposed spectra (Hadrava 2009). These are shown in the  two final columns of Table~1 (after the addition of the system velocity), and agree closely with those obtained from cross-correlation, except around phase 0, where the peaks corresponding to the two components were too closely blended to give reliable results. Essentially, the disentangled spectra of the primary look  the same as those of the combined spectra, whilst those of the secondary are quite noisy owing to the relative faintness of that component (of order one magnitude fainter than the primary).

In {\sc phoebe}, the radial velocities from cross-correlation (excluding those near to phase 0) were used for subsequent modelling, so that a system velocity $\gamma$ could be determined.  The values of mass ratio and $a \sin i$ (with $i$ set to $90^{\circ}$) were also converged as a check on those found by disentangling, giving results of $q=0.20 \pm 0.02$ and $a \sin i=15.1 \pm 0.4~\mathrm{R}_{\odot}$. Since the mass ratios found by each method were significantly different, the value found from disentangling was preferred as being derived from a larger data set and consistent over different spectral regions.  The values of $a \sin i$ and $\gamma$ were then reconverged, with fixed $q$, simultaneously in {\sc phoebe}, giving $15.2 \pm 0.3~\mathrm{R}_{\odot}$ and $-33.9 \pm 1.5$~km~s$^{-1},$ respectively, (uncertainties here are formal errors from the fit to radial velocity curves alone, but are comparable to that found for $a \sin i$ from {\sc korel}).

Following a similar approach to that of Chew (2010), which we also used in Lohr et al. (2014a), after finding the parameters to which radial velocity curves are sensitive, the (unbinned) full SuperWASP photometric data were used to determine the angle of inclination ($i$), the Kopal potentials ($\Omega_1$ and $\Omega_2$), and the temperature ratio of components ($T_1/T_2$), to which light curves are primarily sensitive.  (The very different quantities and qualities of our two main data sources would have skewed our results if simultaneous fitting had been attempted.)  Although a range of different fitting modes were trialled, only Wilson \& Devinney's Mode 5 (semi-detached binary with secondary filling Roche lobe) reproduced both the eclipse depths and the out-of-eclipse variation with any degree of success.  Therefore the secondary Kopal potential ($\Omega_2$) did not need to be determined, since it would be fixed to the value of its Roche lobe.  The primary temperature $T_1$ was also fixed to $7000 \pm 200$~K from our initial spectral fitting (although this had been done using composite spectra, the disentangled spectral regions for the dominant primary component were very similar to those for the combined spectrum).  Other modelling assumptions were: zero eccentricity (since, as noted earlier, the eclipse spacing and durations supported a circular orbit); solar metallicities; surface albedos of 1.0; gravity brightening exponent of 1.0 for the primary (radiative atmosphere) and 0.32 for the secondary (default for convective atmosphere); and a square-root law for limb darkening. The limb darkening parameters were automatically computed by {\sc phoebe} using the $T$ and $\log g$ values to interpolate in tables from van Hamme (1993).

\begin{figure}[ht]
   \begin{center}
   \includegraphics[scale=0.5,angle=0]{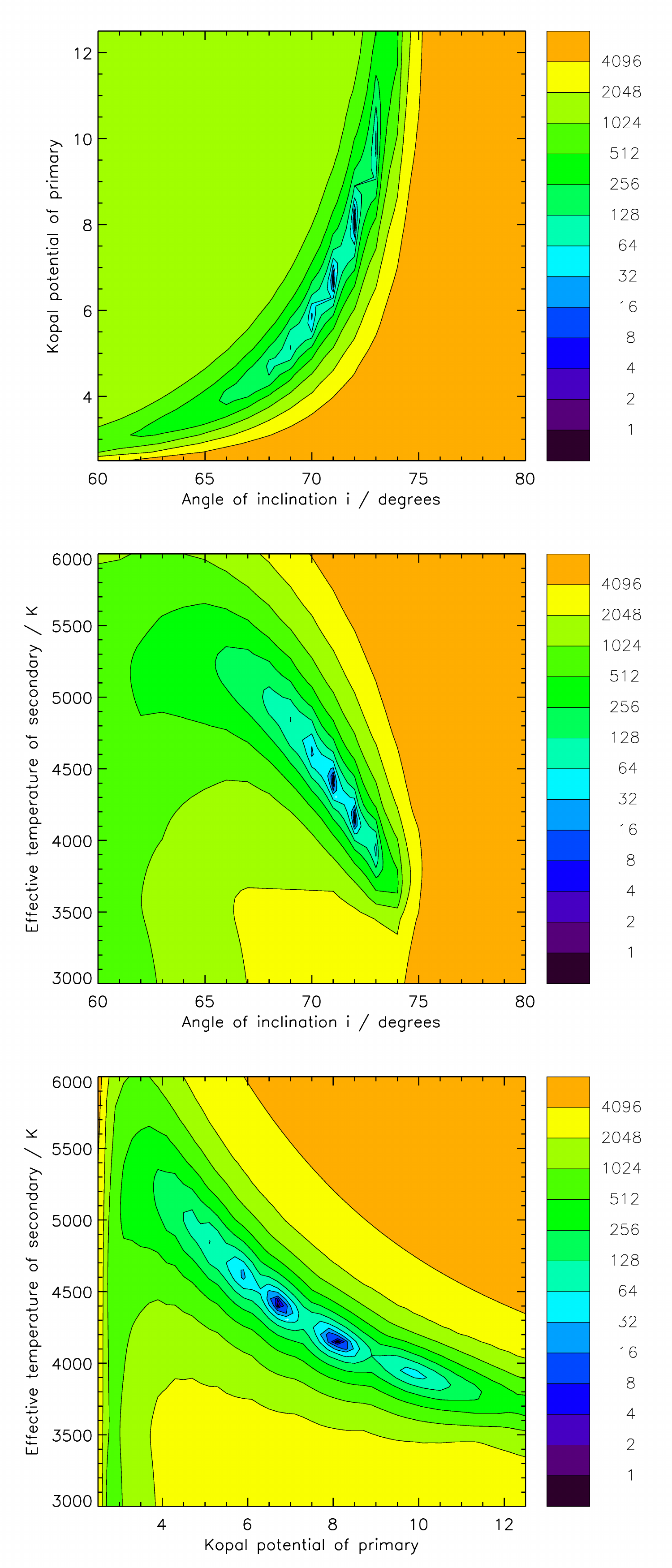}
      \caption{Joint confidence levels for pairs of parameter values ($i$, $T_2,$ and $\Omega_1$ in various combinations) obtained from a series of heuristic scans over a grid around the best fit region. Darker colours correspond to better fits, as indicated by the colour bar; the numerical values represent the number of standard deviations away from the best-fit value.}
   \label{scans}
   \end{center}
   \end{figure}

Using a guideline of Wilson (1994), that the ratio of eclipse depths equals the ratio of eclipsed star surface brightnesses, and that the latter gives a non-linear measure of the components' temperature ratio, $T_2$ was initially set to 4500~K to give a surface brightness ratio comparable to the eclipse depth ratio. Then $i$, $T_2,$ and $\Omega_1$ were varied over a wide-spaced grid of plausible values in a heuristic scan, and the goodness of fit was evaluated for each combination. This search revealed only one region of parameter space that corresponded to acceptable fits, within which a number of local minima are seen, two of which have very similar depths (see Fig.~\ref{scans}). The first of these minima (with $i \sim 72^{\circ}$, $T_2 \sim 4150$~K and $\Omega_1 \sim 8.1$) gives a slightly better fit to the photometric light curve out of eclipse, but a much poorer fit during the primary and secondary eclipses, whereas the second minimum (with $i \sim 71^{\circ}$, $T_2 \sim  4300$~K, and $\Omega_1 \sim 6.7$) gives a better fit to the photometric light curve overall, and so is preferred. To determine the uncertainties on these parameters, we selected 21 sub-samples of the photometric light curve at random (each containing 10000 data points) and found the best fit to each of these sub-datasets in conjunction with the radial velocity data. The distributions of the best fit values for the $i$, $T_2,$ and $\Omega_1$ parameters were then used to assign error bars to each parameter, which represent the standard deviation or one-sigma confidence levels. This, in turn, allowed us to assign confidence levels to the contours in Fig.~\ref{scans}, as shown by the colour bars. Optimal values for $i$, $T_2,$ and $\Omega_1$ were finally obtained by simultaneous convergence within {\sc phoebe}. After each iteration, the value of $a$ was recalculated to maintain $a \sin i$ at the value previously found from the radial velocity curves. This procedure gave values as shown in Table~2; uncertainties are formal errors from the fit, except for those of $i$, $T_2,$ and $\Omega_1,$ which were determined from the procedure described above.

The output values from the modelling (absolute masses, radii, bolometric magnitudes, and surface gravities) are also shown in Table~2.  The uncertainties in the masses were obtained using the formulae prescribed in the {\sc phoebe} manual, and are dominated by the uncertainty in the semi-major axis given by the radial velocities.

The uncertainties in the other output values were found by setting relevant input parameters to their extrema in appropriate combinations, and thus may be expected to overestimate the true uncertainties slightly.  The final fits to radial velocity and light curves, with their residuals, are shown in Figs 7 and~\ref{lcfit}, and images of the best-fitting model at two phases are presented  in Fig.~\ref{schema}, revealing the substantial ellipsoidal distortion of the secondary.

   \begin{figure}[ht]
   \begin{center}
   \includegraphics[scale=0.5,angle=0]{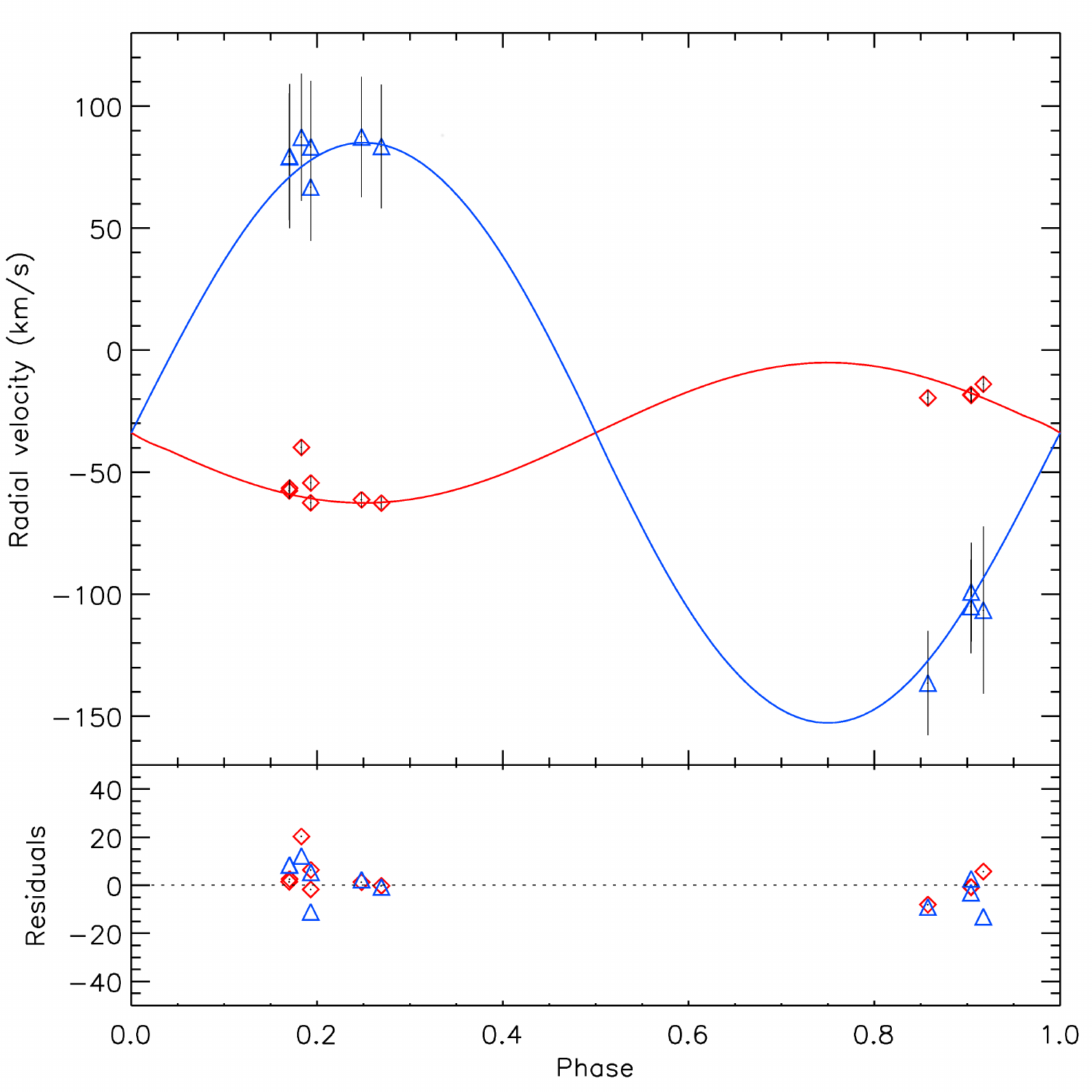}
      \caption{Radial velocity curves of the two components of \object{1SWASP~J050634.16--353648.4}. The best-fitting model from {\sc phoebe} is over-plotted showing the motion of the primary (red) and secondary (blue). The lower panel shows the residuals between the fit and the data, revealing no systematic trends.}
   \label{rvcurve}
   \end{center}
   \end{figure}

\begin{table}
\caption{System parameters for \object{1SWASP~J050634.16--353648.4}}
\begin{tabular}{lll} \hline
Parameter & Value  & Comment \\ \hline
$P_{\mathrm{orb}}$ & 5.104238~d & fixed \\
$T_0$ & HJD 2456992.483978 & \\
$a$ & $16.05 \pm 0.33$~R$_{\odot}$ & \\
$i$ & $71.13^{\circ} \pm 0.05^{\circ}$ & \\
$q$ & $0.236 \pm 0.004$ & \\
$\gamma$ & $-33.9 \pm 1.5$~km~s$^{-1}$ & \\
$e$ & 0 &  assumed \\
$T_1$ & $7000\pm 200$~K & assumed \\
$T_2$ & $4304\pm 9$~K  & given $T_1 = 7000$~K\\
$T_2/T_1$ & $0.6149 \pm 0.0013$ \\
$\Omega_1$ & $6.92 \pm 0.04$ & spherical \\
$\Omega_2$ & $2.17$  & Roche-lobe filling \\
$M_1$ & $1.73 \pm 0.11$~M$_{\odot}$ & \\
$M_2$ & $0.41 \pm 0.03$~M$_{\odot}$ & \\
$R_1$ & $2.40 \pm 0.07$~R$_{\odot}$ & \\
$R_2$ & $4.21 \pm 0.11$~R$_{\odot}$ & \\
$M_{V,1}$ & $2.05 \pm 0.18$ & \\
$M_{V,2}$ & $2.94 \pm 0.19$ & \\
$\log g_1$ & $3.913 \pm 0.016$ & \\
$\log g_2$ & $2.799 \pm 0.007$ & \\ \hline
\end{tabular}
\end{table}

   \begin{figure}[ht]
   \begin{center}
   \includegraphics[scale=0.5,angle=0]{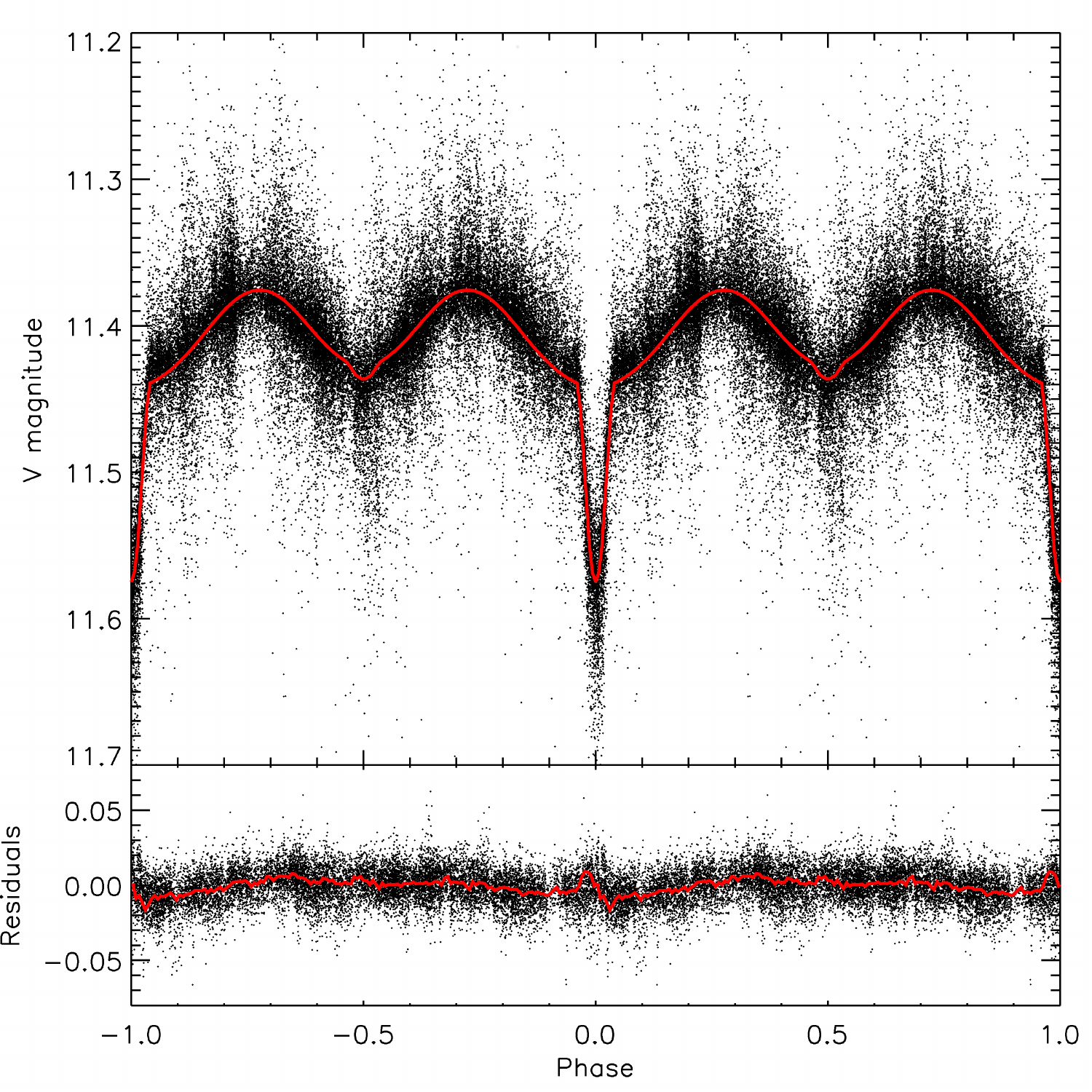}
      \caption{Eclipsing binary light curve for \object{1SWASP~J050634.16--353648.4}, over-plotted with the best-fitting model from {\sc phoebe}. The lower panel shows the residuals between the model fit and the binned data (which we emphasize was not used in the modelling). Around ingress to, and egress from, the primary eclipse, the rapid changes are not as well sampled by the SuperWASP data as in the out-of-eclipse regions, and so give rise to slightly higher residuals of order 0.01 magnitudes.}
   \label{lcfit}
   \end{center}
   \end{figure}

   \begin{figure}[ht]
   \begin{center}
   \includegraphics[scale=0.5,angle=0]{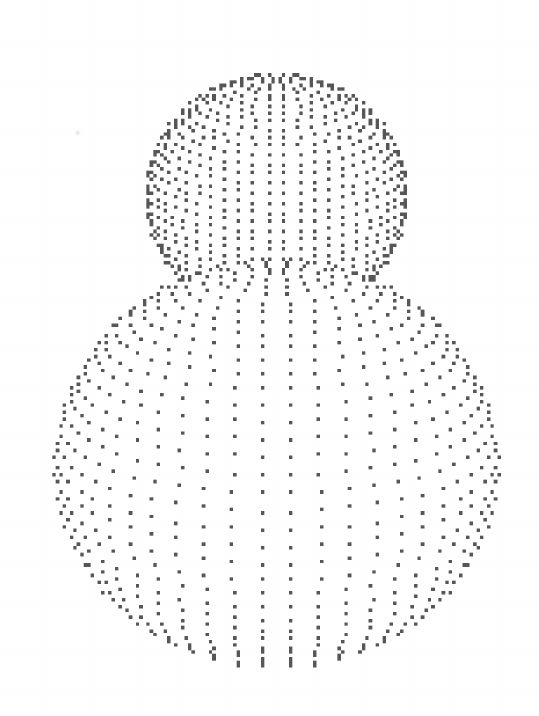}
   \includegraphics[scale=0.5,angle=0]{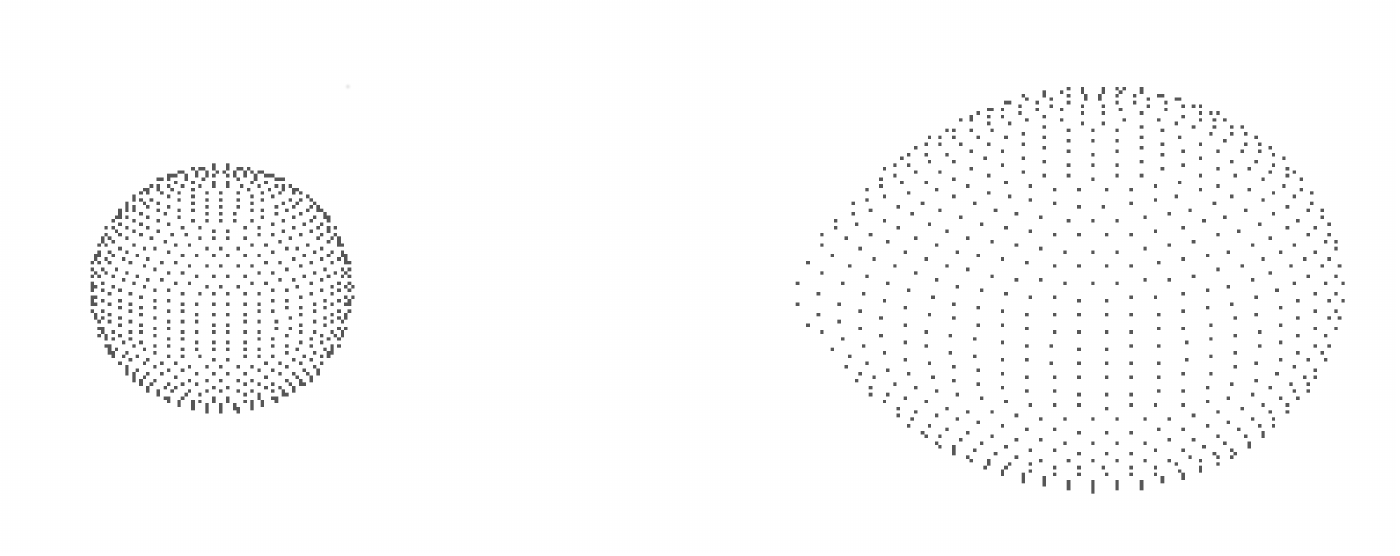}
      \caption{Schematic diagrams of \object{1SWASP~J050634.16--353648.4} at phases 0 and 0.25, according to the best fitting model from {\sc phoebe}.}
   \label{schema}
   \end{center}
   \end{figure}

For comparison with the parameters we have derived, we note that elsewhere \object{1SWASP~J050634.16--353648.4} is included as HE0504--3540 in the catalogue of metal-poor stars from the Hamburg/ESO objective prism survey (Christlieb et al. 2008), where it is listed as having an estimated [Fe/H] value of between --2.1 and --2.9 and a negligible interstellar reddening of $E(B-V) = 0.01$. We note, however, that these  metallicity values may be affected by the composite nature of the spectrum.

 In the fourth data release of the {\sc rave} experiment (Kordopatis et al. 2013), \object{1SWASP~J050634.16--353648.4} is listed as having an effective temperature of 6227~K and $\log g = 1.82$ with a metallicity of --0.06. Since the {\sc rave} spectra are obtained in the infrared (8410 -- 8794 \AA), where the cooler secondary star will make a significant contribution to the overall spectrum, these parameters are clearly composite values for the unresolved binary.

\section{Discussion}

\subsection{Pulsation analysis}

Although the photometric data span over seven years and comprise more than 56000 data points, they are, nonetheless, obtained using 11~cm aperture telescopes from a ground-based site. Consequently, the data do not permit the sort of detailed frequency analyses that have been carried out on similar pulsating stars in eclipsing binaries using data from \textit{Kepler} and \textit{CoRoT} (e.g. Hambleton et al. 2013; da Silva et al. 2014). Nonetheless, we have subtracted the mean orbital modulation from the lightcurve and then performed a {\sc clean} analysis of the resulting detrended data to search for pulsational frequency components. The only frequency components remaining in the power spectrum above the amplitude noise threshold of $\sim 2$~mmag are the fundamental pulsation frequency and its first harmonic (along with $\pm 1$ sidereal day sidebands for each). The amplitude of the first harmonic is $15\%$ that of the fundamental frequency, and the relative phasing between them is such that the fundamental leads the first harmonic by about $60^{\circ}$.

Using the relationship between the pulsation frequency of a given radial mode $f_{\mathrm{pulse}}$, the pulsation constant $Q$ and the ratio of the mean density of the star to that of the Sun
\[ Q f_{\mathrm{pulse}} / (\mathrm{d}^{-1}) = \sqrt{\frac{ \bar{\rho}}{\rho_{\odot}} }, \]
we see that, for the observed fundamental frequency of 13.45~d$^{-1}$, the pulsation constant is $Q=0.026 \pm 0.002$. Alternatively, using the relation from Breger et al. (1993), which is adapted from that in Petersen \& Jorgensen (1972)\[ \log Q = -6.456 + \log (P_{\mathrm{pulse}}/\mathrm{d}) + 0.5 \log g + 0.1 M_{\mathrm{bol}} + \log (T_{\mathrm{e}}/\mathrm{K}), \]
we also calculate $Q=0.026 \pm 0.002$. According to Stellingwerf (1979) this is typical of a first overtone ($n=1$) radial pulsation in a $\delta$~Sct star with the parameters we have modeled here, and similarly Petersen \& Jorgensen (1972) derive a theoretical pulsation constant of $Q_1 = 0.0252$ for the first overtone radial mode, in agreement with our calculated value.

We note the period-luminosity relation for short-period pulsating stars derived by Poretti et al. (2008):
\[ M_V = -1.83(8) - 3.65(7) \log_{10} (P_{\mathrm{pulse}}/{\mathrm{d}}). \]
Using our observed value for the observed pulsation period of 0.0743~d, yields $M_V = 2.29 \pm 0.16$, which is in excellent agreement with the bolometric luminosity derived from the {\sc phoebe} model fit (Table 2), thereby lending further support for the interpretation.

The absence of any pulsation frequencies in the detrended power spectrum at multiples of the orbital frequency indicates that we do not detect any tidally excited modes. However, given that the orbit is apparently circular, this is to be expected.

\subsection{In context with other systems}

The parameters of \object{1SWASP~J050634.16--353648.4} are remarkably similar to those of the star \object{UNSW--V--500} which was announced by Christiansen et al. (2007) as the first high-amplitude $\delta$ Scuti (HADS) star in an eclipsing binary system. That too is a semi-detached system in a 5~d orbit with primary and secondary masses and radii of 1.49~M$_{\odot}$ + 0.33~M$_{\odot}$, and 2.35~R$_{\odot}$ + 4.04~R$_{\odot}$, respectively. Temperatures, surface gravities, and bolometric luminosities are also similar between the two systems, although the pulsational amplitude of UNSW--V--500 is about five times larger than that of \object{1SWASP~J050634.16--353648.4}, as discussed below. Nonetheless, UNSW--V--500 also pulsates in the single first overtone radial mode, with a pulsation constant of $Q=0.025 \pm 0.004$, which is again the same as that of our target.

Liakos \& Niarchos (2015), in an update to Liakos et al. (2012), list 107 objects as eclipsing binaries that contain $\delta$~Sct stars.
Only four of those on the original list from Liakos et al. (2012), namely \object{KW Aur}, \object{TZ Eri}, \object{BO Her,} and \object{MX Pav}, have comparable or larger pulsational semi-amplitudes to our object, being in the range 68 -- 83~mmag. However, \object{1SWASP~J050634.16--353648.4} is not regarded as a HADS star, as such a designation is reserved for those with pulsational semi-amplitudes in excess of 300~mmag. Only two HADS stars in eclipsing binaries are known: the star \object{UNSW-V-500} (Christiansen et al. 2007), with a semi-amplitude of 350~mmag, and \object{RS Gru} (Derekas et al. 2009), with a semi-amplitude of 600~mmag. All HADS stars typically  have dominant pulsations in the fundamental radial mode (McNamara 2000), although some double-mode systems also pulsate in the first and second overtone radial modes. \object{Unusually, UNSW-V-500}, like our target,  is a single-mode field star pulsating only in the first overtone radial mode.

Liakos \& Niarchos (2015) present a convincing empirical correlation between the orbital period and pulsation period of all $\delta$~Sct stars in eclipsing binaries with $P_{\mathrm{orb}} < 13$~d:
\[ \log_{10} (P_\mathrm{pulse}/\mathrm{d}) = 0.56(1) \log_{10} (P_\mathrm{orb} / \mathrm{d}) - 1.52(2). \]
The periods of \object{1SWASP~J050634.16--353648.4}, namely $P_\mathrm{pulse} = 0.0743$~d and $P_\mathrm{orb} = 5.104$~d, fit this relationship perfectly: a pulse period of $(0.075 \pm 0.003)$~d is predicted, based on the orbital period, and an orbital period of $(5.0 \pm 0.4)$~d is predicted, based on the pulse period. Our object thus lends further support to their empirical fit. Although we note that the parameters of two recently described systems with high-quality space-based photometry -- \object{KIC 4544587} (Hambleton et al. 2013) and \object{CoRoT~105906206} (Da Silva et al. 2014) -- do not follow the proposed relation.

Liakos \& Niarchos (2015) also show that the pulse periods of $\delta$~Sct stars in binary systems are correlated with their surface gravity according to
\[ \log g = -0.5(1) \log_{10} (P_{\mathrm{pulse}}/\mathrm{d}) + 3.4(2). \]
Once again, this fits perfectly with \object{1SWASP~J050634.16--353648.4}, predicting $\log g = 4.0 \pm 0.2$ for the pulsating star, in line with that observed.

\subsection{Evolutionary state}

The semi-detached configuration implied by the results of the {\sc phoebe} modelling indicates that mass transfer may be a possibility in \object{1SWASP~J050634.16--353648.4}. However, we see no evidence for current mass transfer in either the photometry or spectroscopy. The heights of the maxima in the binary lightcurve are the same, indicating that there are unlikely to be significant spotted regions on either component. There is no long-term period change detected over the seven year baseline. There are no emission lines seen in the spectrum and there is no X-ray emission seen in the \textit{ROSAT} All Sky Survey that is coincident with the location of this object, the nearest X-ray source being 14 arcmin away. Each of these reasons point to negligible mass transfer currently occurring.

Mkrtichian et al. (2004) introduced the classification of oEA (oscillating EA) binaries for eight systems that are classified as Algol-type eclipsing binaries with mass accreting pulsating components (i.e. \object{Y Cam}, \object{AB Cas}, \object{RZ Cas}, \object{R CMa}, \object{TW Dra}, \object{AS Eri}, \object{RX Hya,} and \object{AB Per}). Amongst the eclipsing binaries with $\delta$ Sct components listed by Liakos \& Niarchos (2015), over half are listed as semi-detached, but it is not clear in what proportion of them mass transfer is currently occurring.

\begin{figure}[ht]
   \begin{center}
   \includegraphics[scale=0.5,angle=0]{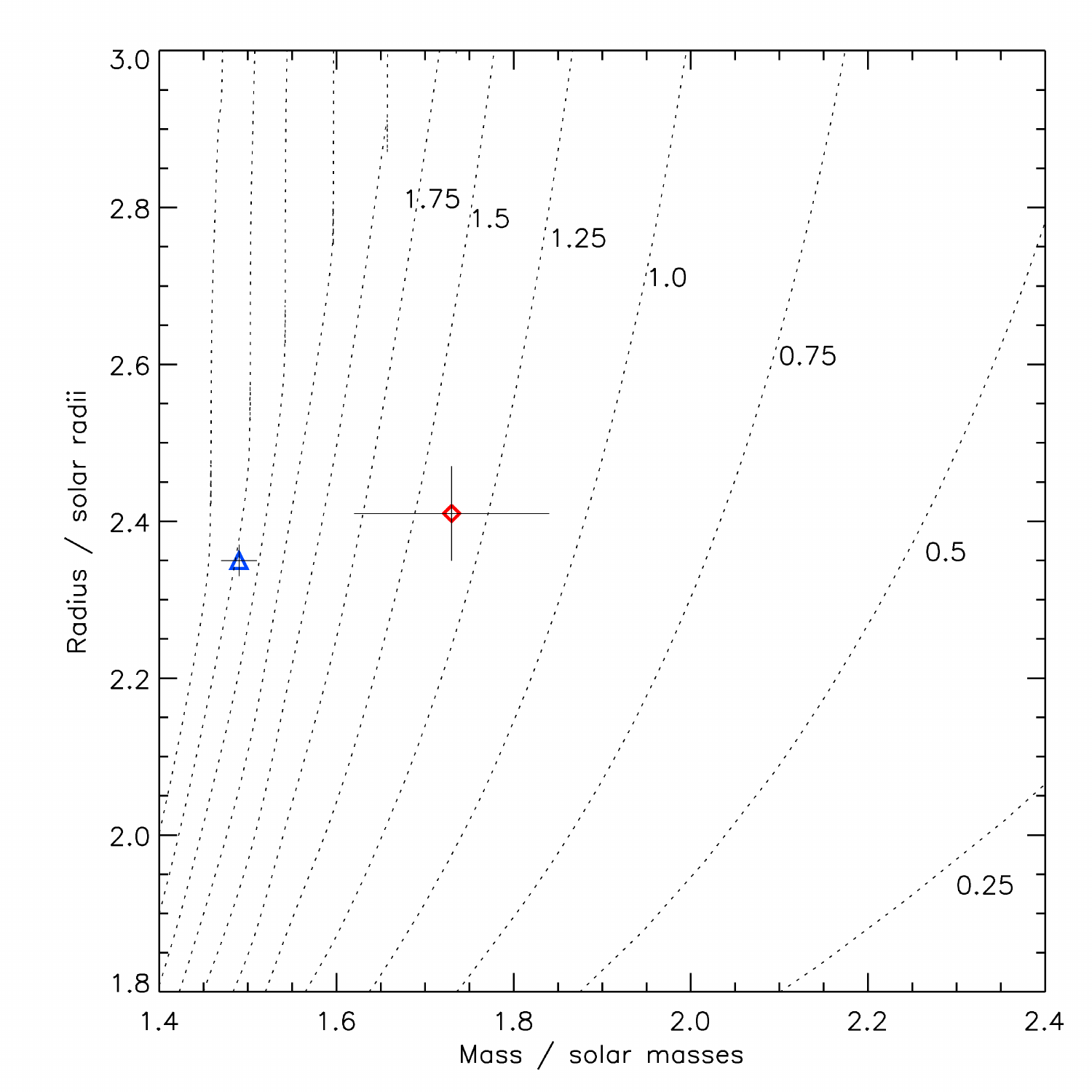}
      \caption{A mass-radius plot showing the positions of \object{1SWASP~J050634.16--353648.4} (red diamond) and \object{UNSW-V-500} (blue triangle) over-plotted on Dartmouth model isochrones, with ages indicated in Gy.}
   \label{massrad}
   \end{center}
   \end{figure}

In Fig.~\ref{massrad} we show a mass-radius plot, which illustrates the position of \object{1SWASP~J050634.16--353648.4,} along with that of \object{UNSW-V-500} for comparison. These are overplotted on Dartmouth model isochrones (Dotter et al., 2008), revealing that the $\delta$~Sct star in our target is somewhat younger than that in \object{UNSW-V-500}, in the range $1.1 - 1.8$~Gy. (The companion, non-pulsating stars, in each system are \emph{not} plotted, since they fill their Roche lobes and so their radii do not represent the size of the corresponding single star.) We also note that, in the presence of possible mass transfer between the two components of each system, the use of single star tracks for the pulsating components may also misrepresent their ages.

\section{Conclusion}

\object{We can confirm that 1SWASP~J050634.16--353648.4} is  a $\delta$~Sct star with a 65~mmag semi-amplitude, 1.784~h pulsation in an eclipsing binary, whose orbital period is 5.104~d. The primary has a mass and radius of 1.7~M$_{\odot}$ and 2.4~R$_{\odot}$ with $\log g = 3.9$ and is around spectral type F1 with an age of $\sim 1.1-1.8$~Gy; the secondary fills its Roche lobe with a mass and radius of  0.4~M$_{\odot}$ and 4.2~R$_{\odot}$ with $\log g = 2.8$. The pulsation constant is $Q=0.026 \pm 0.002$ and indicates a first overtone radial pulsation mode; no tidally excited pulsations are apparent. The system is thus a relatively rare, long period, semi-detached eclipsing binary containing a $\delta$~Sct star with a large pulsation amplitude, similar to \object{UNSW-V-500}.

\begin{acknowledgements}
The WASP project is currently funded and operated by Warwick University and Keele University, and was originally set up by Queen's University Belfast, the Universities of Keele, St. Andrews and Leicester, the Open University, the Isaac Newton Group, the Instituto de Astrofisica de Canarias, the South African Astronomical Observatory and by STFC. Some of the observations reported in this paper were obtained with the Southern African Large Telescope (SALT), under programme 2014-2-SCI-010 (PI: Norton), for which the Open University is a shareholder, as part of the UK SALT Consortium. One of us (Lohr) is supported under the STFC consolidated grant to the Open University, ST/L000776/1. The research  made use of the SIMBAD database, operated at CDS, Strasbourg, France.
\end{acknowledgements}


\end{document}